\documentclass[a4paper]{jpconf}
\usepackage{graphicx}

\begin{document}
\title{Wave-particle duality in classical mechanics}

\author{Alexander Y Davydov}

\address{AlgoTerra LLC, 249 Rollins Avenue \# 202, Rockville, MD 20852, USA}

\ead{alex.davydov@algoterra.com}

\begin{abstract}Until recently, wave-particle duality has been thought of as quantum principle without a counterpart in classical physics. This belief was challenged after (i) finding that average dynamics of a classical particle in strong inhomogeneous oscillating field resembles that of a quantum object and (ii) experimental discovery of "walkers" - macroscopic droplets that bounce on a vertically vibrating bath of the same fluid and can self-propell via interaction with the surface waves they generate. This paper exposes a new family of objects that can display both particle and wave features all together while strictly obeying laws of the Newtonian mechanics. In contrast to the previously known duality examples in classical physics, oscillating field or constant inflow of energy are not required for their existence. These objects behave deterministically provided that all their degrees of freedom are known to an observer. If, however, some degrees of freedom are unknown, observer can describe such objects only {\it probabilistically} and they manifest {\it weird} features similar to that of quantum particles. We show new classical counterparts of such quantum phenomena as particle interference, tunneling, above-barrier reflection, trapping on top of a barrier, and spontaneous emission of radiation. In the light of these findings, we hypothesize that quantum mechanics may \emph{emerge} as approximation from a more profound theory on a deeper level.\end{abstract}

\section{Introduction}
Wave-particle duality is one of the fundamental principles of quantum mechanics which is directly linked to many of its mysteries. It postulates a dual nature of particles and radiation in microworld by enforcing the trade-off between the particle-like and the wave-like behavior of a quantum object, depending on experimental conditions. Since its original formulation as a hypothesis by Louis-Victor de Broglie in 1923 [1], the duality principle has been supported by numerous experiments involving photons [2], electrons [3-6], neutrons [7], atoms and dimers [8], small van der Waals clusters [9] and, more recently, $C_{60}$ fullerenes [10, 11]. Perhaps the most famous among these is the double-slit experiment, in which, let say, electrons are shot at a barrier with two slits, travel through the slits and are detected on a wall behind the barrier. The interference pattern formed by the electrons on the wall strongly suggests that they behave like waves in this setup. The interference pattern remains even when the electrons are shot separately, with interval between shots being large enough to ensure that only a single electron is present in the system at a time. If, however, an experimentator decides to determine through which slit each electron passes and sets up a detector close to one of the slits, mysteriously electrons stop creating an interference pattern, i.e., they start behaving like classical particles.            

Until quite recently, it was widely believed that wave-particle duality has peculiarly quantum nature, an attribute of behavior of matter at small length and energy scales that does not have a counterpart in classical physics. This belief remained unshaken by some known examples of 'classical tunneling' - crossings of potential barriers by objects of insufficient kinetic energy under condition that these objects are extended in space so that not all their parts are crossing the barrier region simultaneously (e.g., the Fosbury flop in high jump or a long train passing over the short hill) [12, 13]. The situation changed when in 2005 I. Dodin and N. Fisch showed that the average dynamics of a classical particle in a high-frequency oscillating field resembles motion of a quantum particle in a conservative field, with particle average displacement during the oscillation period playing the role of an effective de Broglie wavelength $\lambda$. In a quasiclassical field, with a spatial scale large compared to $\lambda$, the guiding-center motion is adiabatic. Otherwise, a particle exhibits quantized eigenstates in ponderomotive potential wells, tunnels through "classically forbidden" regions, and experiences stochastic reflection from attractive potentials [14-16]. In another development at approximately the same time, Y. Couder {\it et. al} discovered that macroscopic oil droplets can bounce on the vertically oscillating liquid bath and interact with the surface waves they generate to form stable self-propelled symbiotic structures called "walkers" [17, 18]. Walkers were shown experimentally to exhibit interference in double-slit experiment [19] along with other quantum-like phenomena such as unpredictable tunneling [20] and orbit quantization in a rotating frame [21]. 

The goal of this paper is to further extend the list of classical examples of wave-particle duality by presenting a rather general mechanism by which such examples can be produced in the framework of purely classical physics. Herein we present new classical counterparts of such quantum phenomena as particle interference, tunneling, above-barrier reflection, trapping on top of a potential barrier, and spontaneous emission of radiation. This list is likely to be continued increasing the number of links between classical and quantum theories. These links deserve serious attention and would be good topics for future research. Nevertheless, based on what is known so far including the findings presented in this paper, one cannot help but wonder whether the relationship between quantum and classical physics is much closer than is currently acknowledged. Especially intriguing is the question of whether quantum mechanics is an {\it incomplete} theory that emerges from the underlying classical sub-dynamics. This question has been the subject of long and on-going debates among physicists starting from the birth of quantum theory and has lead to many interesting theoretical developments including the stochastic mechanics introduced by E. Nelson [22, 23], statistical mechanics of matrix models of S. Adler [24], and emergent quantum mechanics based on nonequilibrium thermodynamics on a sub-quantum level pioneered by G. Gr\"ossing [25].                   

\section{Classical mechanics of soft bodies}
Let us define a \emph{soft body} (SB) as a collection of identical point particles (\emph{monads}) of mass {\it m} coupling pairwisely with attractive central force. One can think of SB as a cloud consisting of interacting monads. Monads are confined to the cloud as a whole since parting the cloud into distant peaces would require a significant amount of energy. However, the cloud's overall shape can easily vary with time, thus distinguishing a soft body from a rigid one. The dynamics of monads inside SB is determined by coupling between them as well as by external forces that monads can also 'feel'. For simplicity sake, we assume that monad-monad interaction is given by an attractive harmonic potential, i.e., the force exerted upon a monad at location ${\textbf{r}}_{i}$ by another monad at location $\textbf{r}_{j}$ is proportional to a distance $\mid\textbf{r}_{i}-\textbf{r}_{j}\mid$. It follows that monads can penetrate through each other effortlessly (cf. \textit{asymptotic freedom}) and, therefore, internal dynamics of monads is collisionless and can be treated analytically. Another assumption is concerned with how an external force acts on a monad. Let $U(\textbf{r},t)$ denote the external potential and $\textit{N}$ be a total number of monads in SB. We assume that a monad located at ${\textbf{r}}_{i}$ effectively 'experiences' the external potential $U(\textbf{r}_{i},t)/N$. This calibration condition ensures that, when all monads happen to be co-located at the same point $\textbf{r}$, the total external potential of SB equals $U(\textbf{r},t)$. Thus, total potential energy of SB can be written as
\begin{equation}
U_{tot}(\textbf{r}_{1},\textbf{r}_{2},\ldots,\textbf{r}_{N})=\frac{1}{4}\:m\omega_{0}^{2}\sum_{all\: pairs}^{\frac{1}{2}N(N-1)}\mid\textbf{r}_{i}-\textbf{r}_{j}\mid^{2}\: + \;\frac{1}{N}\sum_{i=1}^{N}U(\textbf{r}_{i},t),
\end{equation}
\

where $\omega_{0}$ denotes the characteristic angular frequency of oscillations of SB. In what follows, we will be concerned with dynamics of the \emph{center of mass} of SB moving in a smooth external potential in the non-relativistic limit. The main question is: How can we describe this motion effectively using as few degrees of freedom as possible?

\subsection{The simplest case: SB comprised from two monads in one dimension}

Consider the simplest possible instance of SB - two monads in one dimension. The total mass of SB reads $M=2m$. Let $x_{1}$ and $x_{2}$ denote the coordinates of two monads.
Assume that the external potential is time-independent: $U(x,t)\equiv U(x)$. Also assume that, at any time instance, a linear size of SB is much smaller than the characteristic length at which $U(x)$ changes substantially. It is convenient to introduce new variables as follows
\begin{equation}
X=\frac{1}{2}\:(x_{1}+x_{2}),\ \xi=\frac{1}{2}\:(x_{1}-x_{2})
\end{equation}
\

where $X$ is a position of the center of mass (CM) and $\xi$ is an internal degree of freedom of SB. Expanding the potential $U(x)$ in power series around CM, we obtain for the total potential energy $U_{tot}$ (up to the higher order terms)
\begin{equation}
U_{tot}(x_{1},x_{2})=\frac{1}{2}\:U(x_{1})+\frac{1}{2}\:U(x_{2})+\frac{1}{4}\:m\omega_{0}^{2}(x_{1}-x_{2})^{2}\cong U(X)+\frac{1}{2}\:U^{''}(X)\xi^{2}+\frac{1}{2}\:M\omega_{0}^{2}\xi^{2},
\end{equation}
\

where prime denotes differentiation with respect to $X$. The total kinetic energy $T_{tot}$ of SB expressed via new variables reads
\begin{equation}
T_{tot}=\frac{1}{2}\: M\dot{X}^{2}+\frac{1}{2}\: M\dot{\xi}^{2},
\end{equation}
\

where the overdot means differentiation with respect to time. From the action integral 
\begin{equation}
S=\int_{t_{1}}^{t_{2}}(T_{tot}-U_{tot})dt\cong
\int_{t_{1}}^{t_{2}}[\:\frac{1}{2}\: M\dot{X}^{2}+\frac{1}{2}\: M\dot{\xi}^{2} - U(X) - \frac{1}{2}\:M(\omega_{0}^{2}+\frac{1}{M}U^{''}(X))\xi^{2}\:]dt
\end{equation}
\
\
we can then write down the Euler-Lagrange equations [26] as conditions for the action $S$ to be stationary, by the Hamilton's principle, as follows
\begin{equation}
M\ddot{X}=-U^{'}(X)-\frac{1}{2}\:U^{'''}(X)\xi^{2}
\end{equation}
\begin{equation}
\ddot{\xi}+[\omega_{0}^{2}+\frac{1}{M}U^{''}(X)]\xi=0
\end{equation}
\
\

Notice rather remarkable relationship between the Eqs. (6) and (7). Equation (6) expresses the Newton's second law for the motion of CM, with the last term on the r.h.s. describing influence of the internal degree of freedom on this motion. Equation (7) can be easily identified as the time-dependent harmonic oscillator equation (TDHOE) for the internal degree of freedom $\xi$, with time dependence entering the picture via the term $U^{''}[X(t)]$. Thus, we see that CM \emph{parametrically} affects oscillations of the internal degree of freedom, which in its turn, can directly influence the motion of CM (in the presence of position-dependent external potential). Also notice the mathematical equivalence of TDHOE to the time-independent Schr\"odinger equation, the root of interesting links between quantum and classical physics [27].

\subsection{Conservation of energy}
When an external potential of SB is (explicitly) time-independent, the total energy is a conserved quantity. In this case, by combining (3) and (4), we obtain 
\begin{equation}
E=\frac{1}{2}\:M\dot{X}^{2}\:+\:U(X)+\frac{1}{2}\:M\dot{\xi}^{2}\:+\:\frac{1}{2}\:M\omega_{0}^{2}\xi^{2}\:+\:\frac{1}{2}\:U^{''}(X)\xi^{2} =constant
\end{equation}
\

The first two terms in (8) give contribution that can be associated with the kinetic and potential energies of CM, the third and fourth terms add up to the 'internal energy', while the fifth term describes the coupling between (in this case) two degrees of freedom of SB and can result in energy exchange between them. Let us emphasize that, unless $\xi\equiv0$ or the external potential $U$ is a constant or a linear function of $X$, the sum $\frac{1}{2}\:M\dot{X}^{2}\:+\:U(X)$ does not conserve during motion. This is the main cause of quantum-like behavior of the CM of SB under various circumstances, as will be clear later.  

\subsection{Time-dependent harmonic oscillator}
It is convenient to rewrite Eq.(7) as follows
\begin{equation}
\ddot{\xi}+\Omega^{2}(t)\xi=0
\end{equation}
\

with $\Omega^{2}(t)\equiv\omega_{0}^{2}+\frac{1}{M}\:U^{''}[X(t)]$. The time-dependent parameter $\Omega(t)$ is allowed to take not only real but pure imaginary values as well, so that $\Omega^{2}(t)$ may become negative for some time intervals without any contradiction [27]. Two linearly independent solutions of Eq.(9) can be written in the polar form
\begin{equation}
\xi_{1}=\rho(t)cos[\theta(t)]\:;\:\xi_{2}=\rho(t)sin[\theta(t)]
\end{equation}
\
and combined into one complex-valued function $\varphi$ as follows
\begin{equation}
\varphi(t)=\xi_{1}(t)+i\xi_{2}(t)=\rho(t)e^{i\theta(t)}
\end{equation}
\

After substituting (11) into Eq.(10) and separately equating real and imaginary terms on both sides, we obtain two equations
\begin{equation}
\frac{1}{\rho}\:\frac{d}{dt}(\rho^{2}\dot{\theta})=0
\end{equation}
and
\begin{equation}
\ddot{\rho}+\Omega^{2}(t)\rho-\rho\:{\dot{\theta}}^{2}=0
\end{equation}
\

Equation (12) can be thought of as the angular momentum conservation law (for two-dimensional motion of a point particle of mass $m$ in the plane ($\xi_{1}$,$\xi_{2})$) and recast as
\begin{equation}
\frac{1}{2}M\rho^{2}\dot{\theta}=J=constant,
\end{equation}
\

where constant $J$ has a dimension of \emph{action} (or angular momentum) and plays an essential role in the theory to be developed below. Expressing $\dot{\theta}$ via $\rho$ from (14) and substituting it into (13), we obtain the celebrated \emph{Ermakov-Pinney} equation [28, 29]
\begin{equation}
\ddot{\rho}+\Omega^{2}(t)\rho-\frac{4J^{2}}{M^{2}\rho^{3}}=0
\end{equation}
\

Alternatively, one can assume that $\dot{\theta}>0$ for all times and, from (14), find $\rho=\sqrt{2J/(M\dot{\theta})}$ and consequently 
$\ddot{\rho}=-\sqrt{\frac{J}{2M}}\:(\dot{\ddot{\theta}}\cdot\dot{\theta}-\frac{3}{2}\ddot{\theta}^{2})/{\dot{\theta}^{5/2}}$. After substituting these results into (13), one finds that the phase function $\theta(t)$ must satisfy the equation
\begin{equation}
\Omega^{2}(t)=\dot{\theta}^{2}+\frac{1}{2}\:\{\theta;t\},
\end{equation}
\ 

where $\{\theta;t\}$ denotes the \emph{Schwarzian derivative} [30] of $\theta(t)$ defined as $\{\theta;t\}\equiv\dot{\ddot{\theta}}/\dot{\theta}-\frac{3}{2}(\ddot{\theta}/\dot{\theta})^{2}$. Hence the (non-conserved) sum of kinetic and potential energies of the time-dependent harmonic oscillator, which coordinate is given by $\xi_{1}(t)$, can be written in the form
\begin{equation}
\epsilon_{1}(t)=\frac{1}{2}\:M\dot{\xi}_{1}^{2}\:+\:\frac{1}{2}\:M\Omega^{2}(t)\xi_{1}^{2}=J\omega_{_{\texttt{eff}}}(t),
\end{equation}
\ 

where the \emph{effective} angular frequency $\omega_{_{\texttt{eff}}}(t)$ is defined as
\begin{equation}
\omega_{_{\texttt{eff}}}(t)\equiv\dot{\theta}+\frac{1}{2}\:cos^{2}(\theta)\cdot(\dot{\ddot{\theta}}\cdot\dot{\theta}-\ddot{\theta}^{2})/{\dot{\theta}^{3}}+\frac{1}{2}sin(2\:\theta)\cdot\ddot{\theta}/\dot{\theta}\:,
\end{equation}
\ 

with $\theta(t)$ being the solution of Eq.(16) for a given $\Omega^{2}(t)$.

The formulae (17) and (18) are general and therefore must be applicable in the particular case when $\Omega^{2}(t)\equiv\omega^{2}_{0}=constant$. In this case, the general solution of Eq.(16) is given by

\begin{equation}
\theta(t)=\arctan{[\kappa\cdot \tan{(\omega_{0}t+\alpha_{0})}]}
\end{equation}
\

where $\kappa$ is an arbitrary positive number, and $\alpha_{0}$ denotes the phase constant. The total energy of the oscillator is, of course, time-invariant in this case and reads
\begin{equation}
\epsilon_{1}=J\omega_{0}/\kappa
\end{equation}

In the light of (17) and (18), we can rewrite the energy conservation law for SB (8) as follows
\begin{equation}
E=\frac{1}{2}\:M\dot{X}^{2}\:+\:U(X)+J\omega_{_{\texttt{eff}}}(t) =constant
\end{equation}

\subsection{Equation of motion for center of mass}
Assuming that $\omega^{2}_{0}\gg\left|U^{\:''}\!(X)/M\right|$ for any $X$, we can employ the Wentzel-Kramers-Brillouin (WKB) method [31] to obtain the approximate solution of Eq.(7):    
\begin{equation}
\xi(t)\approx\frac{\sqrt{2J/M}}{\sqrt[4]{\omega^{2}_{0}+U^{''}[X(t)]/M}}\cdot \cos{\left(\int_{t_{0}}^{t} \sqrt{\omega^{2}_{0}+U^{''}[X(\tau)]/M} \:d\tau + \alpha\right)}
\end{equation}
\

where $\alpha$ is a phase constant, and we have used the condition that, in absence of the external potential (for free SB), $\xi(t)=\sqrt{2J/(M\omega_{0})}\cos{(\omega_{0}t+\tilde{\alpha})}$ with $\tilde{\alpha}\equiv\alpha-\omega_{0}t_{0}$. Notice the dependence of $\xi(t)$ not only upon the value of $X$
at moment $t$ but also upon the whole path history $X(\tau)$ at earlier times $\tau<t$ via the integral in the argument of the cosine function in (22). Interestingly, this feature is reminiscent of "path memory" of walkers, which is believed to be responsible for quantization of their orbits in a rotating bath [21]. Substituting (22) into (6), we finally obtain the integro-differential equation of motion for a CM of SB: 
\begin{equation}
M\ddot{X}=-U^{'}(X)-\frac{J\:U^{'''}(X)}{M\sqrt{\omega^{2}_{0}+U^{''}(X)/M}}\cdot \cos^{2}{\left(\int_{t_{0}}^{t} \sqrt{\omega^{2}_{0}+U^{''}[X(\tau)]/M} \:d\tau + \alpha\right)}
\end{equation}
\

In case $J=0$, i.e., when the internal degree of freedom of SB is not excited, Eq.(23) coincides with the second Newton's law for motion of a point particle of mass \textit{M} in the external potential $U(X)$. However, if $J\neq0$, the behavior of CM can be significantly altered in comparison with that of a point particle of the same mass, presenting a great surprise to an observer who is unaware of that difference and treats SB as a point particle. It is worth emphasizing that Eq.(23) is valid even for a wider class of potentials that explicitly depend on time $U(X,t)$, while the energy conservation law (expressed by (8) or (21)) does not hold in this case.

To further simplify our analysis, henceforth we accept a crude approximation of (23) given by  
\begin{equation}
M\ddot{X}=-U^{'}_{\texttt{eff}}(X,t)
\end{equation}
\begin{equation}
U_{\texttt{eff}}(X,t)\equiv U(X)+\frac{J\cdot U^{''}(X)}{M\omega_{0}}\cdot \cos^{2}{(\omega_{0}t+\alpha)}
\end{equation}
\

where the phase constant $\alpha$ has been redefined to absorb $-\omega_{0}t_{0}$.

\section{Classical examples of wave-particle duality and analogues of quantum mechanical phenomena} 
Here we present new examples of wave-particle duality in the realm of classical mechanics based on the theory developed in section 2.

\subsection{Classical analogue of tunneling}
Consider a potential barrier of a Gaussian shape $U(X)=A\exp\left(\frac{(X-X_{0})^{2}}{2\sigma^{2}}\right)$. Assume that a SB of mass $M$ approaches this barrier from the left with CM moving at constant velocity $V$ such that the corresponding kinetic energy $KE=\frac{1}{2}MV^{2}$ (far from the barrier) is smaller than the  barrier height $A$, as shown in Fig.1. Analysis of Eqs. (24) and (25) reveals that such SB may either reflect back or go through the barrier, depending on the value of the phase $\theta=\omega_{0}t+\alpha$ at moment when the CM is near the barrier top. Figure 2 shows the CM trajectories for different values of phase constant $\alpha$ chosen at random from the continuous uniform distribution on the interval [0,$2\pi$[. Thus, $\alpha$ plays the role of a \emph{hidden variable} that completely determines the result of the interaction of SB with the potential barrier. For an observer who treats this SB as a point particle with the equivalent mass, its behavior seems unpredictable with apparent violation of the energy conservation law for some short time intervals and, therefore, can be classified as \emph{tunneling}.  

\begin{figure}[h]
\begin{minipage}{18pc}
\includegraphics[width=18pc]{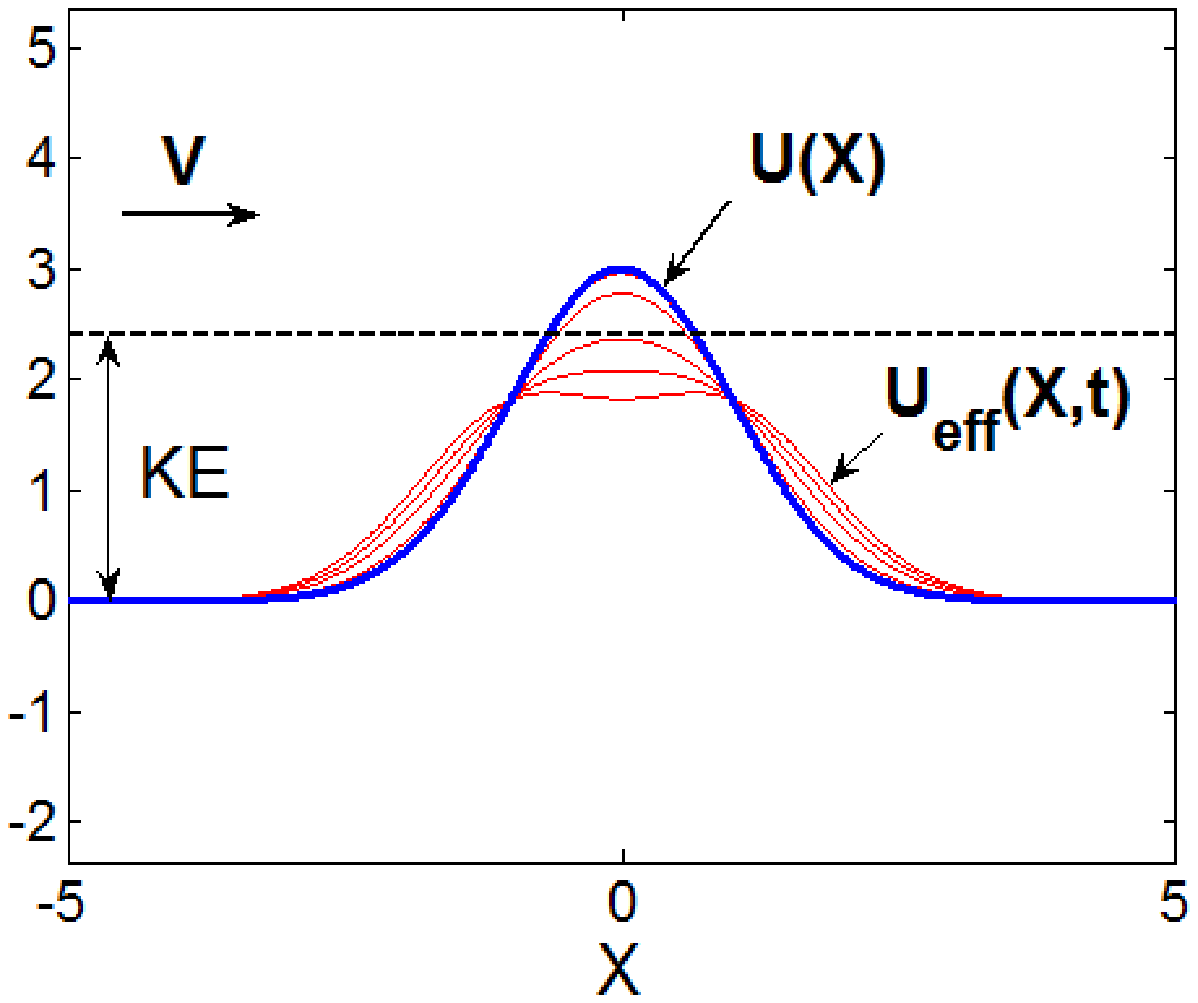}
\caption{\label{label}Setup for classical analogue of quantum tunneling. See text for details.}
\end{minipage}\hspace{1pc}%
\begin{minipage}{18pc}
\includegraphics[width=18pc]{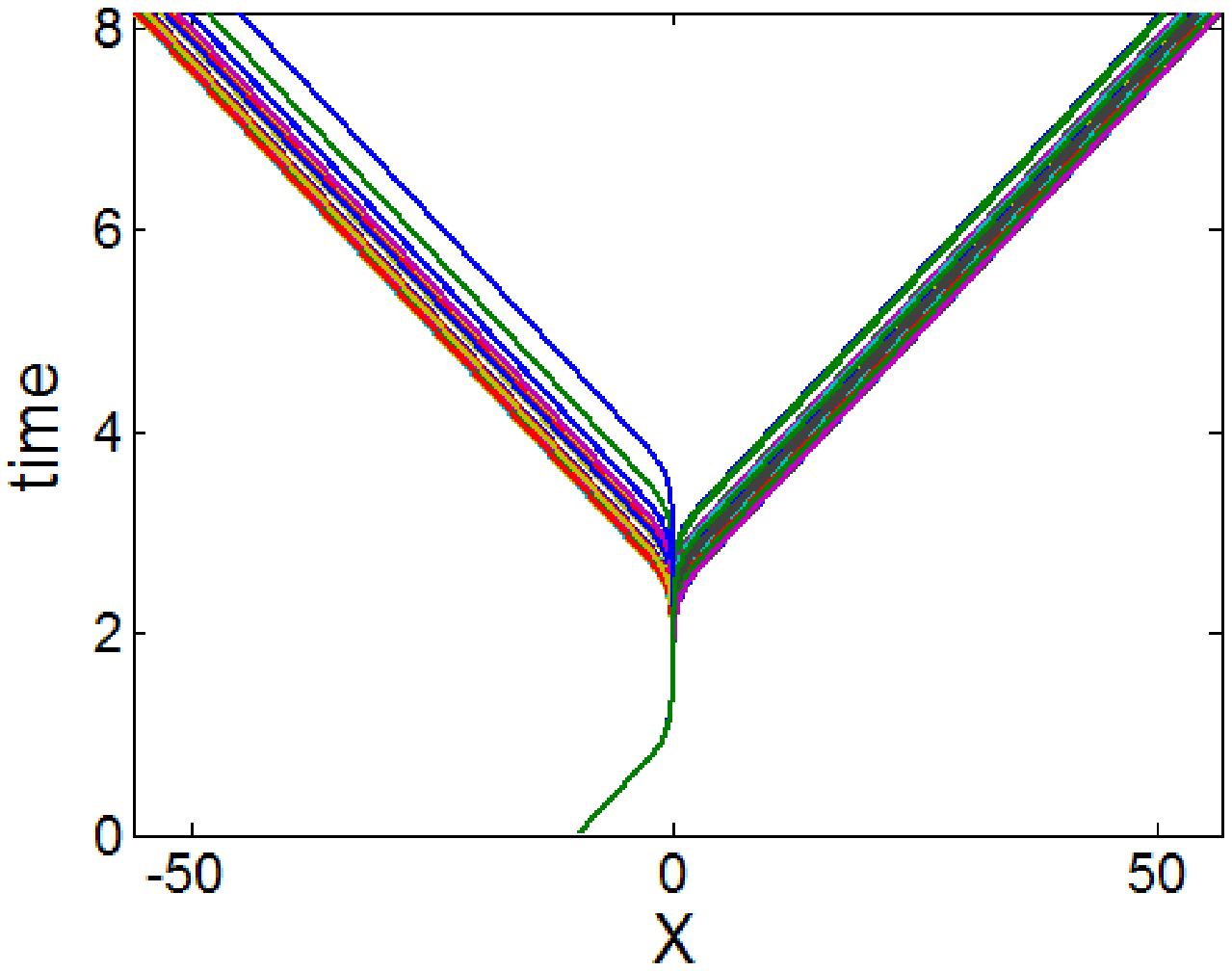}
\caption{\label{label}Trajectories for 100 different values of phase constant $\alpha$.}
\end{minipage} 
\end{figure}

\subsection{Classical analogue of above-barrier reflection}
Consider now the case when $KE>A$ as depicted in Fig.3. With appropriate choice of parameters, it is possible to obtain above-barrier reflection of a SB as shown in Fig.4.

\begin{figure}[h]
\begin{minipage}{18pc}
\includegraphics[width=18pc]{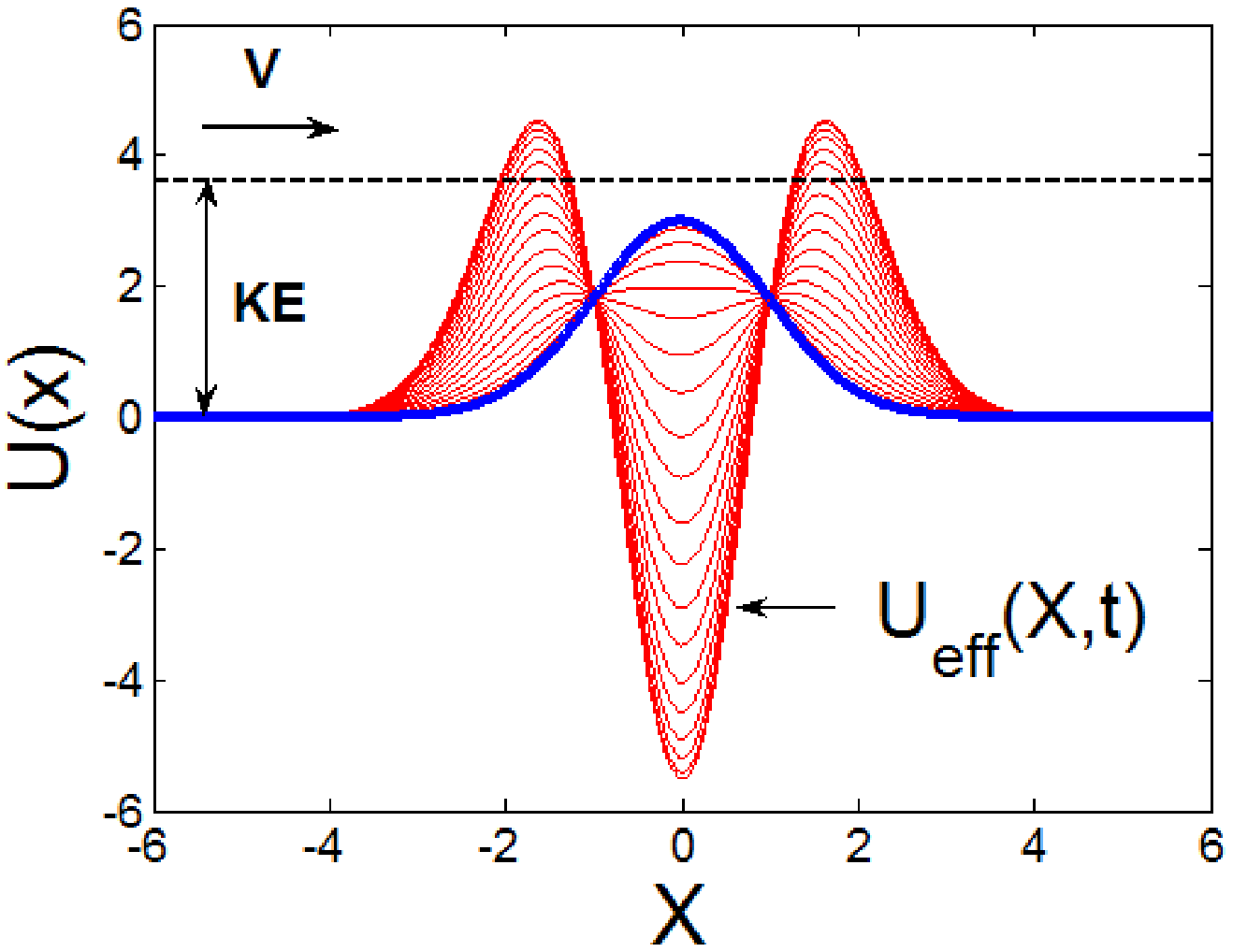}
\caption{\label{label}Setup for classical analogue of above-barrier reflection. See text for details.}
\end{minipage}\hspace{1pc}%
\begin{minipage}{18pc}
\includegraphics[width=18pc]{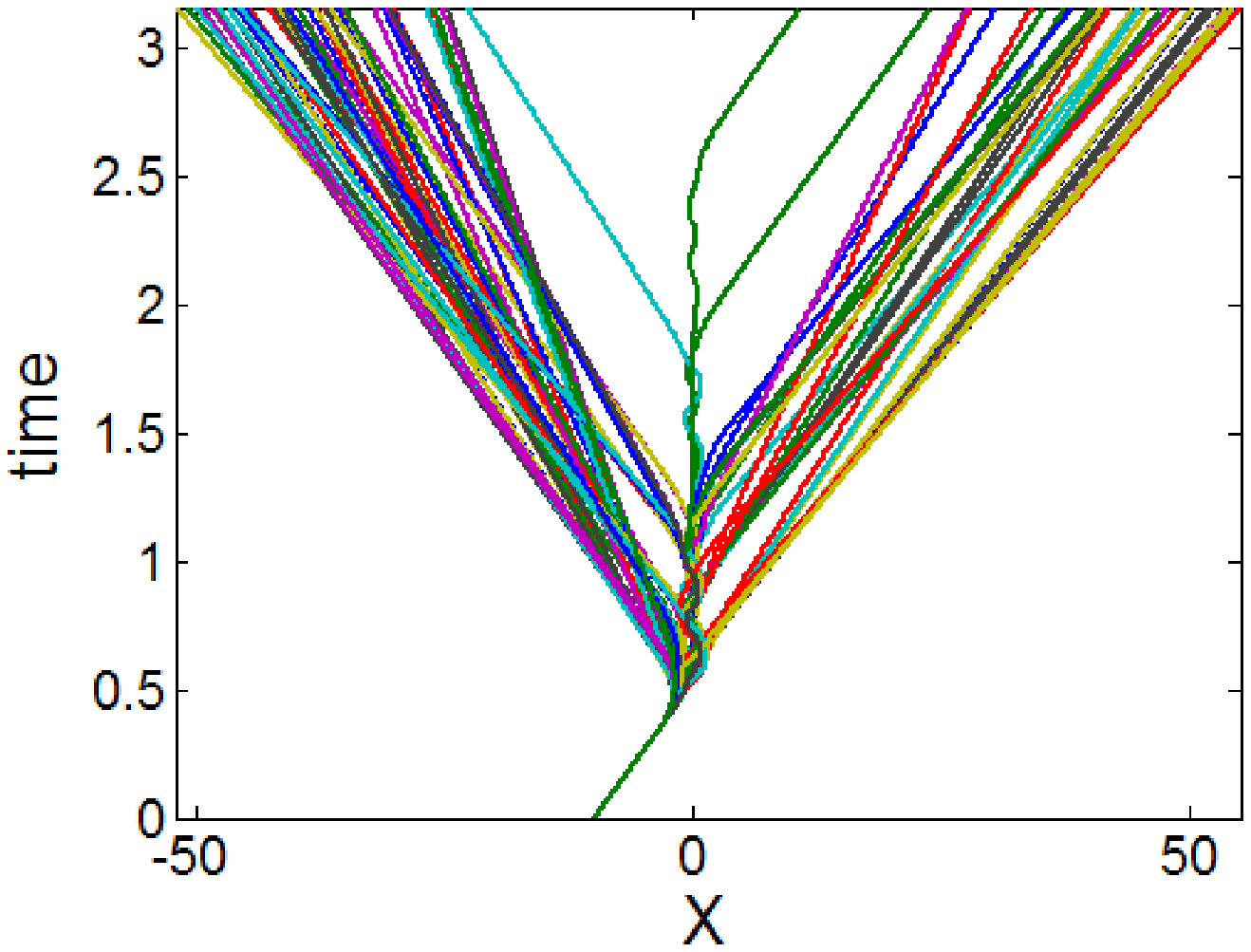}
\caption{\label{label}Trajectories for 100 different values of phase constant $\alpha$ illustrating classical above-barrier reflection.}
\end{minipage}
\end{figure}

\subsection{Classical analogue of particle trapping on top of a barrier and spontaneous emission}
Another example of paradoxical behavior of SB can be obtained if we select $KE>A$ but $KE-A\ll A$, see Fig.5. In this case, apart from the above-barrier reflection, we observe a new phenomenon: the CM can be confined for considerable interval oscillating around the top of the potential barrier and then be released in either direction with equal probability and approximately the same velocity as it had before the confinement. This behavior is similar to that of a quantum particle confined in a potential plateau region, as has been discussed in detail in [32]. Numerical simulations show that direction of the \emph{emission} of a trapped SB is highly sensitive to the phase $\theta$ of the internal degree of freedom, i.e., small change in the value of $\theta$ can lead to emission in the opposite direction.      

\begin{figure}[h]
\begin{minipage}{18pc}
\includegraphics[width=18pc]{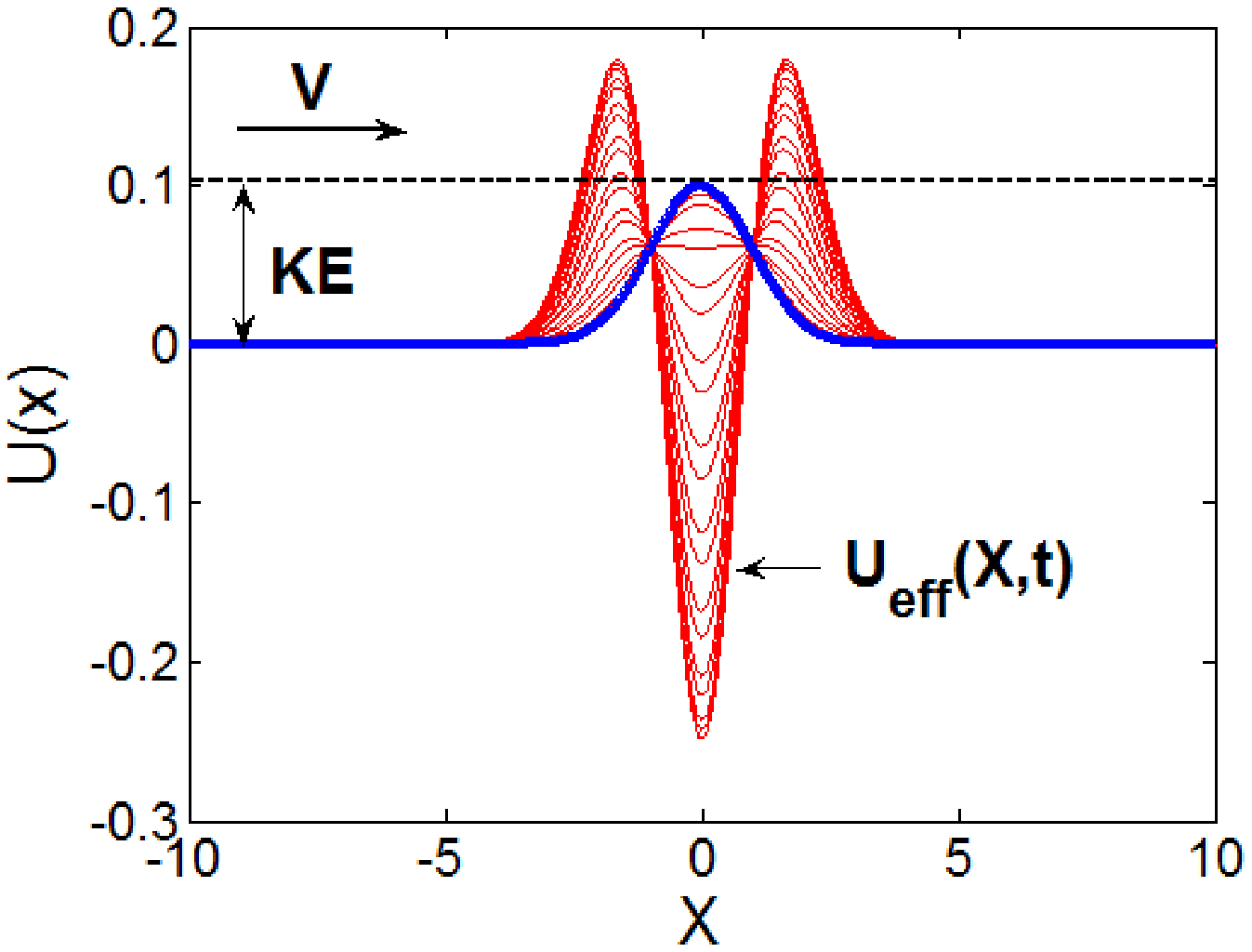}
\caption{\label{label}Setup for classical analogues of paradoxical confinement of a particle on top of a potential barrier and spontaneous emission. See text for details.}
\end{minipage}\hspace{1pc}%
\begin{minipage}{18pc}
\includegraphics[width=18pc]{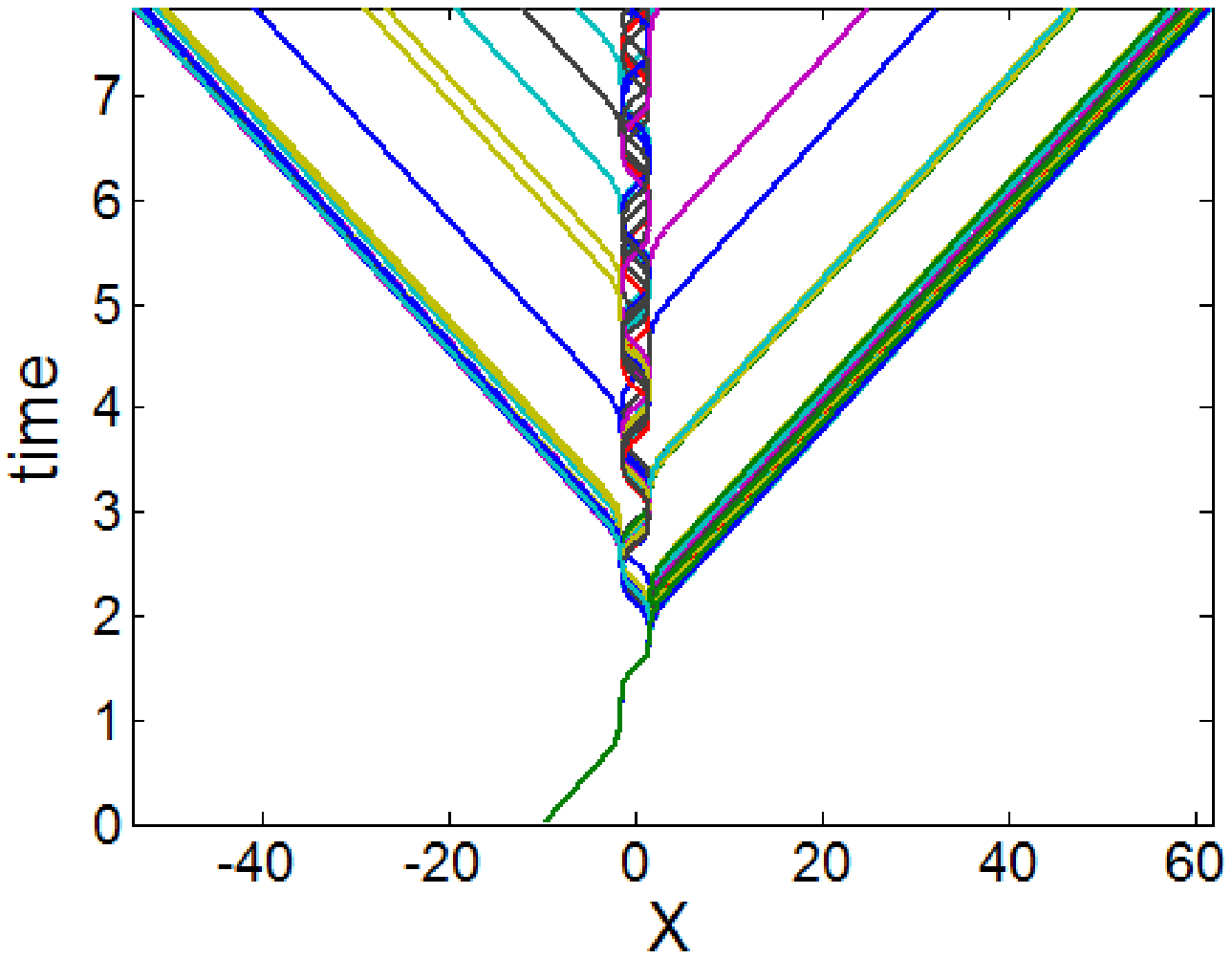}
\caption{\label{label}Trajectories for 100 different values of phase constant $\alpha$ illustrating classical confinement on top of a barrier and subsequent spontaneous emission.}
\end{minipage}
\end{figure}

\subsection{Interference in 1D}
Interference phenomena are ubiquitous in Nature and are most easily understood using the theory of waves. For instance, colorful patterns produced by soap bubbles can be explained by constructive and destructive interference of light waves. It is much more demanding to explain thin-film interference using the representation of light as a collection of photons, requiring concepts of quantum theory, such as probability amplitude, to be invoked. Interestingly, Sir Isaac Newton - who considered light as being composed of tiny particles which he called 'corpuscles' - made some ingenious arguments in attempt to interpret the results of an experiment measuring the relationship between the thickness of a sheet of glass and partial reflection of light off it. In the words of R. Feynman [33], "...to account for the fact that the thickness of the glass determines the amount of partial reflection, Newton proposed this idea: Light striking the first surface sets off a kind of wave or field that travels along with the light and predisposes it to reflect or not reflect off the second surface. He called this process "fits of easy reflection or easy transmission" that occur in cycles, depending on the thickness of the glass".

\begin{figure}[h]
\includegraphics[width=18pc]{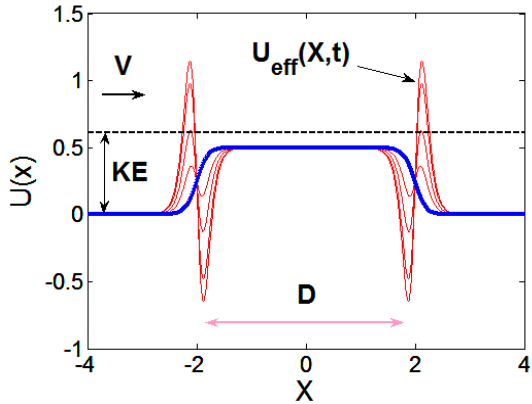}\hspace{2pc}
\begin{minipage}[b]{18pc}\caption{\label{label}Scattering of a SB off a 'soft rectangular' potential barrier. See text for details.}
\end{minipage}
\end{figure}

\begin{figure}[h]
\begin{minipage}{18pc}
\includegraphics[width=18pc]{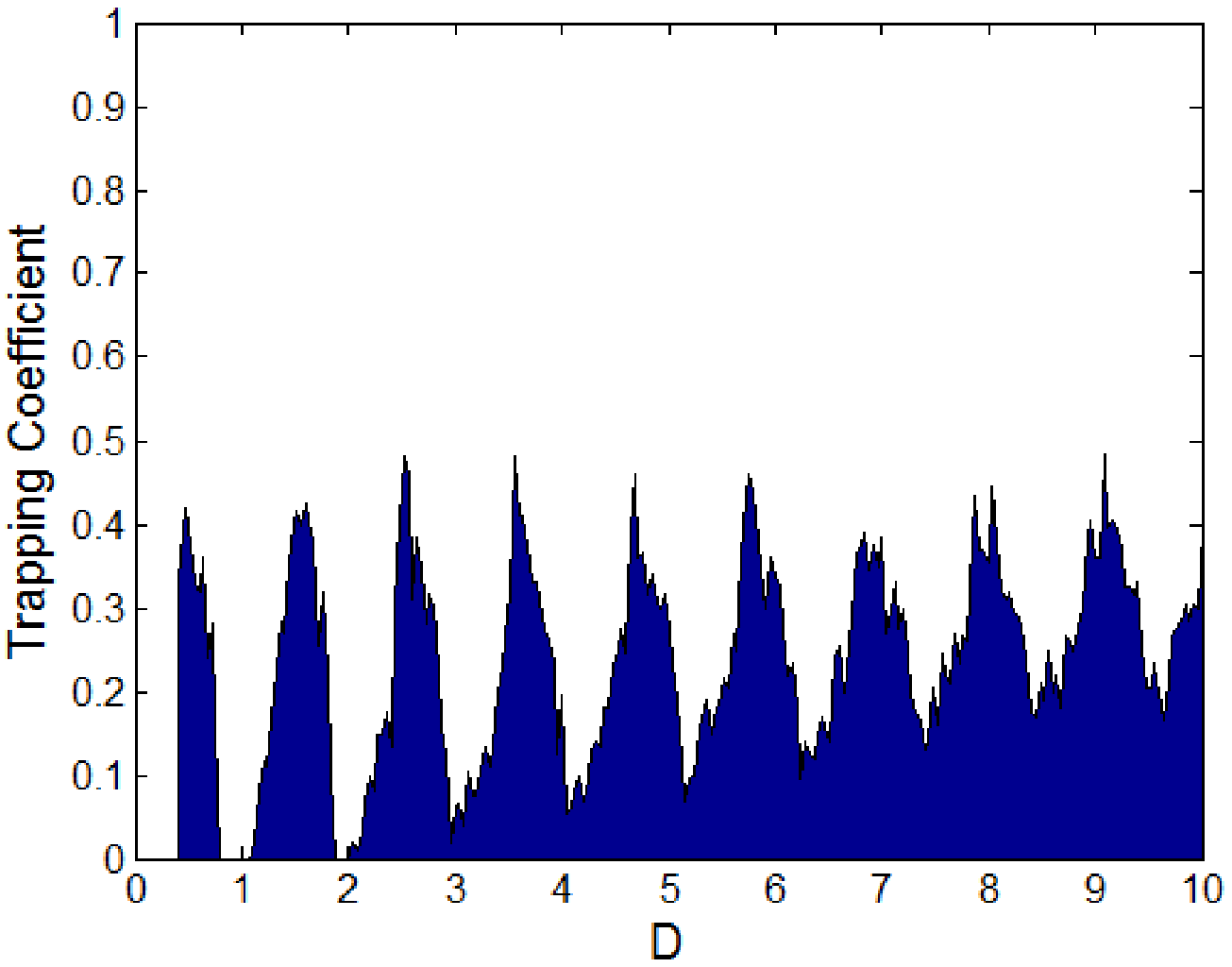}
\caption{\label{label}Trapping coefficient as a function of barrier width $D$ estimated numerically after fixed time interval since the first collision of a SB with the barrier.}
\end{minipage}\hspace{1pc}%
\begin{minipage}{18pc}
\includegraphics[width=18pc]{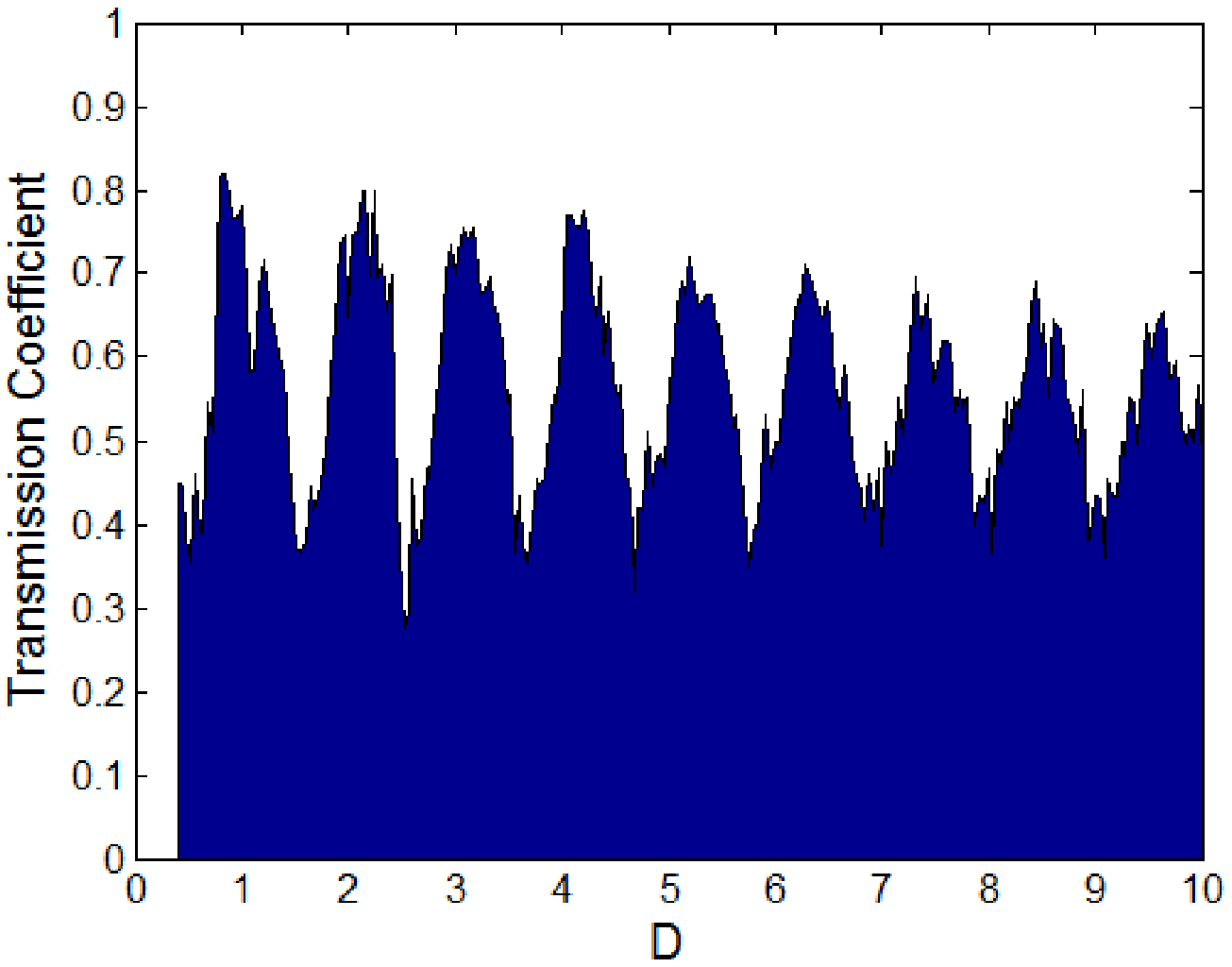}
\caption{\label{label}Transmission coefficient as a function of barrier width $D$ estimated numerically after fixed time interval since the first collision of a SB with the barrier.}
\end{minipage}
\end{figure}

Under specific circumstances, a SB can exhibit behavior alike interference of waves in thin films, although it can be revealed only \emph{statistically} as a build-up process of an interference-like pattern from single events and, besides, has a temporal character. Consider an ensemble of soft bodies of the same type prepared such that all members have fixed kinetic energy $KE$ of their CM but different phase constants $\alpha_{i}$ drawn from the continuous uniform distribution on the interval $[0,2\pi[$. Each SB moves from the left toward the potential barrier of 'soft rectangular' shape $U(X)=\frac{A}{2}\left( \tanh\left(\frac{X+d}{L}\right)-\tanh\left(\frac{X-d}{L}\right) \right)$ shown in Fig. 7. The width of the barrier is given by $D\equiv2d$. The kinetic energy of CM exceeds the barrier height: $KE>A$. We define the fraction of soft bodies in the ensemble trapped inside the barrier at some time instance $t=\tau_{m}$ as the \textit{trapping coefficient}. The \textit{transmission coefficient} is defined along the similar lines. Results of numerical simulations demonstrate that both trapping and transmission coefficients, estimated at some fixed $\tau_{m}$, depend quasi-periodically upon the width of barrier $D$, as shown in Figures 8 and 9. This oscillatory dependence fades away with increasing $\tau_{m}$, as more and more soft bodies of the ensemble which are temporally confined inside the barrier start to leak out in either back or forward direction until none are left. 

\section{Conclusions} 
We have presented a new family of classical objects (soft bodies) obeying laws of the Newtonian mechanics that can exhibit particle and wave properties simultaneously. Such systems extend the list of known examples of the wave-particle duality in classical mechanics. They do not require an oscillating external field or constant inflow of energy to manifest their dual properties and behave deterministically provided that all degrees of freedom are known to an observer. If some degrees of freedom are unknown, soft bodies can be described only {\it probabilistically} and  manifest {\it weird} features similar to that of quantum particles including interference, tunneling, above-barrier reflection, trapping on top of a barrier,  and spontaneous emission of radiation. Further developments are underway. In the light of these findings, it is tempting to hypothesize that quantum mechanics may \emph{emerge} as approximation from a more profound theory on a deeper level. If true, the road to achieving complete understanding of quantum theory may lie through an attempt to construct an \textit{average dynamics} of classical objects like soft bodies, in which unobserved degrees of freedom have been effectively integrated out. This average dynamics will necessarily have a probabilistic nature. One can speculate that a concept of \textit{wave function} may emerge as a mathematical tool to cope with lack of information about all degrees of freedom of a soft body, and the Schr\"odinger equation may even be \textit{deduced} from the first principles. Such program is in line with the vision of A. Einstein who predicted: "Assuming the success of efforts to accomplish a complete physical description, the statistical quantum theory would, within the framework of future physics, take an approximately analogous position to the statistical mechanics within the framework of classical mechanics. I am rather firmly convinced that the development of theoretical physics will be of this type, but the path will be lengthy and difficult." [34]. The present paper advocates making new steps along this path.     

\section*{Acknowledgements} 
I thank Andrej Ahmeteli for interesting discussions and bringing the Ref. [13] to my attention. Also I am grateful to Ilya Dodin for providing me with references to his work on quantum-like behavior of classical particles in oscillating fields.

\smallskip

\end{document}